\documentstyle[aps,multicol,pre]{revtex}

\title{On the double criticality of fluids adsorbed in disordered porous 
media}
\author{E.V. Vakarin$^a$, W. Dong$^b$  and J.P. Badiali$^a$}
\address{$^a$ UMR 7575 LECA ENSCP-UPMC, 11 rue P. et M. Curie, 75231 Cedex
05, Paris, France\\
$^b$ Laboratoire de Chimie, UMR 5182 CNRS, Ecole Normale
Sup\'{e}rieure de Lyon, 46 All\'{e}e
d'Italie, 69364 Lyon, Cedex 07, France}

\begin{document}
\maketitle
\begin{abstract}
The phase transition of a fluid adsorbed in a heterogeneous system is 
studied with two simple lattice gas models within the framework of a 
mean-field theory. Despite the different origin of the heterogeneity 
(spatial variation of binding energy or fluid coordination numbers), the 
fluid can undergo two phase transitions if the hosting system is 
sufficiently heterogeneous. It is clearly shown that such a 
polymorphism in the confined fluid results from the successive 
condensations in distinct spatial regions of the host. We have found 
the precise conditions at which  one two phase transitions occur. The 
insight gained from the present study allows one to understand better some 
recent puzzling simulation results. 
\end{abstract}
%\begin{document}
%\maketitle

%\begin{multicols}{2}

\section{Introduction}

It is well known that fluids exhibit
a behavior quite different from that in the bulk when adsorbed in
porous matrices (see Ref.~\cite{review}
for a recent review). The main features are the lowering of the critical
temperature and shrinking of the liquid-gas coexistence region.
The effect of a porous medium on the phase diagram
has recently been the subject of intensive theoretical studies,
aiming at
analyzing the relevant microscopic mechanisms. The matrix confinement and
disorder are thought to be the key factors, which have been
studied\cite{comment} by means of computer
simulations\cite{Monson1,Monson3,Levesque,Wei1,Wei2} and analytical
models\cite{RFIM,Kierlik1,Kierlik2,KierlikOZ,Shapir2,Sarkisov,Sok}.
In addition, the pore size characterization indicates\cite{Gubbins}
a significant impact of the pore blocking and connectivity effects.

Despite a significant progress in our understanding of all these
processes, there remain some controversial points.  One of them is
the existence of the second critical point which is believed to signal the
wetting\cite{Monson1} (or "precondensation"\cite{Kierlik1}) transition.
The second phase transition has been detected in a series of
independent computer simulation
studies\cite{Monson1,Monson3,Levesque,Wei1} of Lennard-Jones
fluids in a variety of matrix models. The location of the critical
point has been
found to be quite structure-dependent\cite{Monson3,Levesque,Wei1}. Moreover,
two matrix realizations of the same porosity led to qualitatively different
criticalities\cite{Wei1} (either one or two transitions). Sarkisov
and Monson\cite{Monson3} even questioned the very existence of the second
critical point, showing that it can disappear after the averaging over
several matrix realizations.

Analytical models have not given a definite answer. Kierlik {\it et
al} \cite{Kierlik1} studied the model of lattice gas with
quenched impurities (LGQI). A mean-field approximation resulted in the usual
criticality, while the inclusion of correlation effects (in the
mean-spherical approximation) yielded the second transition. Unfortunately,
it was not possible to draw a clear conclusion because for three-dimensional
lattices the same approach gives the second transition even for
the bulk fluid.
A similar dilemma resulted from the theoretical approximations has been reported\cite{KierlikOZ}
for an off-lattice model studied within the Ornstein-Zernike approach.
Recent rigorous results for LGQI model on a Bethe lattice\cite{Sok} did not
give any evidence for the second transition (at least in the range of
parameters the authors have studied).

Our main objective is to find out the criterion discriminating
these
two types of behavior. For this purpose, we analyze  a lattice gas
model in the mean-field approximation.  This  analytically tractable
model incorporates all the key features listed above.
For this reason, one may expect that such an approach can help to understand better the double criticality.

%%%%%%%%%%%%%%%%%%%%%%%%%%%%%%%%%
\section{Role of disorder and porosity}
%%%%%%%%%%%%%%%%%%%%%%%%%%%%%%

\subsection{Random binding lattice gas}

%%%%%%%%%%%%%%%%%%%%%%%%%%%%%%%%%

Let us consider a lattice of $N$ sites with the coordination
number equal to $q$.
Each site can be occupied by fluid or matrix particles with the
corresponding
occupation numbers $t_i=0$ (fluid particle absent at site i), $t_i=1$ 
(fluid particle present at site i), $x_i=0$ (matrix particle present at site 
i) and $x_i=1$ (matrix particle absent at site i). If the number of matrix 
sites is $N_m$, then the "nominal" number of the fluid-accessible sites is 
$N_f=N-N_m$ and the overall porosity is $\Phi=(1-N_m)/N=N_f/N$. The matrix 
configuration $\{x_i\}$ is quenched. Different choices  for the set 
$\{x_i\}$ correspond to different matrix realizations.
The Hamiltonian is given by

\begin{equation}
H=-W\sum_{<ij>}t_it_j -K\sum_{<ij>}t_i(1-x_j),
\end{equation}
where $W$ is the fluid-fluid interaction constant and $K$ describes
the fluid-matrix interaction.
The first double sum is over the fluid-accessible nearest neighbors
($x_i=x_j=1$). The second summation runs over the fluid-matrix nearest
neighbors.
Such a version of the lattice gas with quenched impurities
\cite{Kierlik1,Kierlik2,Shapir2} is studied in ref.~\cite{Sok}.
In the limit of very high porosity, $N_f \approx N$  and
$\sum_{i=1}^{N_f} \approx \sum_{i=1}^{N}$.
Note that the term
$K\sum_j(1-x_j)$ can be interpreted as a random energy parameter 
$E_i$ associated with the site $i$. Such a random binding arises from
the random environment of adsorption sites in the presence of the matrix. 
Therefore, the effect of matrix can be considered to produce simply the random binding through the pore space (sites occupied by fluid particles).
In this way, we arrive at the random binding lattice gas model with the following Hamiltonian,
\begin{equation}
\label{Hamiltonian}
H=-W\sum_{<ij>}t_it_j - \sum_i E_i t_i.
\end{equation}
By introducing the spin-like variables $s_i=2t_i-1$, the above
model reduces readily to the random field Ising model(RFIM)\cite{RFIM}.

In order to study how the phase behavior depends on the averaging over
the sample realizations, we do not focus on a specific matrix
realization but consider an ensemble of matrix samples from which we can
extract a probability distribution $f(E_i)$ for the binding energies.
For simplicity, we choose a non-correlated multi-modal form,
\begin{equation}
f(E_i)=\sum_{a}c_{a}\delta(E_i-\epsilon_{a}),
\end{equation}
where $c_{a}$ is the concentration of sites with the binding energy
$\epsilon_{a}$.
In terms of fluctuation $t_i=\theta +(t_i-\theta)$, the product of occupation numbers can be rewritten as,
\begin{equation}
\label{fluctuation} 
t_it_j=\theta^2+\theta(t_i-\theta)+\theta(t_j-\theta)+
(t_i-\theta)(t_j-\theta),
\end{equation}
where $\theta$ is the mean fluid density whose expression will be
given shortly later (see eq.~(\ref{coverage})).
Neglecting the product of fluctuations, {\it i.e.}, the last term on the right hand side of eq.~(\ref{fluctuation}), we obtain,
\begin{equation}
\label{MFA}
t_it_j=t_i\theta+t_j\theta-\theta^2.
\end{equation}
Substituting eq.~(\ref{MFA}) into eq.~(\ref{Hamiltonian}) leads finally to the Hamiltonian of the mean field approximation (MFA).
The corresponding grand partition function for a particular matrix
configuration is given by,
\begin{equation}
\Xi(\{E_i\})=
\sum_{\{t_i\}} e^{-\beta H}e^{\beta \mu \sum_{i}^{N_f}t_i}=
e^{-\beta qW\theta^2 N_f/2}\prod_{i=1}^{N_f}\left[
1+e^{\beta(\mu+E_i+qW\theta)} \right],
\end{equation}
where $\beta=1/(kT)$ ($T$: temperature and $k$: Boltzmann constant). 
In the following, we consider
attractive fluid-fluid interactions ({\it i.e.},$W>0$).

All the thermodynamic quantities can be obtained after averaging over 
the quenched disorder. The pressure is given by
\begin{equation}
\label{pressure}
P=\frac{kT}{N_f} \int \prod_{i=1}^{N_f} dE_i f(E_i) 
\ln\left[\Xi(\{E_i\})\right]=
- qW\theta^2/2 + kT\sum_{a} c_a \ln\left[
1+e^{\beta(\mu+\epsilon_a+qW\theta)}\right].
\end{equation}
For a particular disorder configuration, the fluid density varies from site to site. The density at site j is given by,
\begin{equation}
<t_j(E_j)>=
\frac{1}{\Xi(\{E_i\})}\sum_{\{t_i\}} t_j e^{-\beta H}e^{\beta \mu \sum_{i}^{N_f}t_i}=
\frac{e^{\beta(\mu+E_j+qW\theta)}}{1+e^{\beta(\mu+E_j+qW\theta)}}.
\end{equation}
Averaging over the disorder, we obtain the mean fluid density,
\begin{equation}
\label{coverage}
\theta=
\int dE_j f(E_j) <t_j(E_j)> =
\sum_{a}c_{a}\frac{e^{\beta(\mu+\epsilon_{a}+qW\theta)}}
{1+e^{\beta(\mu+\epsilon_a+qW\theta)}}.
\end{equation} 
It is easy to check that the above MFA results are thermodynamically 
consistent, {\it i.e.}, $P$ and $\theta$ given in eqs.~(\ref{pressure}) and 
(\ref{coverage}) satisfy the Gibbs-Duhem relation, 
\begin{equation}
\label{GibbsDuhem}
\left(\frac{\partial P}{\partial \mu}
\right)_T=\theta.
\end{equation} 

Considering a bimodal distribution with the binding energies $\epsilon_0$ and
$\epsilon_1=\epsilon_0+\Delta$ and concentrations $c_0$ and $c_1=1-c_0$ 
we arrive at
\begin{equation}
\label{bcoverage}
\theta=c_0\tau_0+c_1\tau_1,
\end{equation}
where
$$
\tau_0=\frac{e^{\beta(\mu+\epsilon_0+qW\theta)}}
{1+e^{\beta(\mu+\epsilon_0+qW\theta)}};
\qquad
\tau_1=\frac{e^{\beta(\mu+\epsilon_0+\Delta+qW\theta)}}
{1+e^{\beta(\mu+\epsilon_0+\Delta+qW\theta)}}
$$
are the local densities in the regions with different
binding energies.
Solving eq.~(\ref{coverage}) with respect to $\mu$ for the bimodal case, we 
obtain 
\begin{equation}
\label{mu}
\mu=-\epsilon_0-qW\theta+kT\ln \left[
\frac{G-\sqrt{G^2-4\theta R(\theta-1)}}
{2R(\theta-1)}
\right],
\end{equation}
where
$$
G=-\theta(1+R)+R(1-c_0)+c_0; \qquad R=e^{\beta\Delta}.
$$
In the limit of vanishing heterogeneity $\Delta=0, R=1$ we recover
the conventional mean-field result for a bulk fluid,
$$
\mu=-\epsilon_0-qW\theta+kT\ln \left[\frac{\theta}{1-\theta}\right].
$$
The compressibility can be found in the standard way,
$$
\chi^{-1}=\theta^2\left(\frac{\partial \mu}{\partial \theta}\right)_T.
$$
Solving $\chi^{-1}=0$ with respect to $T$, we can determine the 
spinodal curve.
The result for the spinodal is analytical but
requires a very heavy algebra. Moreover, the criticality
conditions,
\begin{equation}
\label{conditions}
\left(\frac{\partial \mu}{\partial \theta}\right)_T=0, \qquad
\left(\frac{\partial^2 \mu}{\partial \theta^2}\right)_T=0,
\end{equation}
lead to equations for $T_c$ and $\theta_c$ which are not
analytically soluble. For this reason, we introduce some further
approximations which allow us to study more explicitly the critical 
point in a few limiting cases.

In the case of weak heterogeneity (small $\Delta$ or $R-1$), we can expand
$\mu$ in eq.~(\ref{mu}) to the second order in $(R-1)$ and obtain,
\begin{equation}
\mu=-\epsilon_0-qW\theta+kT\left[\ln \left(\frac{\theta}{1-\theta}\right)
+(c_0-1)(R-1) + \Psi(c_0,\theta)(R-1)^2\right],
\end{equation}
where
$\Psi(c_0,\theta)=\frac{1}{2}-c_0(1-\theta)-c_0^2(\theta-\frac{1}{2})$. From
the criticality conditions (\ref{conditions}), we get
\begin{equation}
\label{criticalT}
\frac{T_c}{T_c^{bulk}}=1-\frac{c_0(1-c_0)}{4}
\left[
e^{\Delta/kT_c} -1
\right]^2;
\qquad
\theta_c=1/2,
\end{equation}
where 
$T_{c}^{bulk}=qW/4k$ is the bulk critical temperature.
To the second order of (R-1), the result for the critical density, 
{\it i.e.}, $\theta_c=1/2$, is identical to that for a bulk fluid.
However, eq.~(\ref{criticalT}) shows clearly that the critical temperature 
is always lower than that of the bulk fluid, {\it i.e.},  
$T_c < T_c^{bulk}$. The lowering of critical temperature of confined 
fluids has been observed in computer simulations 
\cite{Monson1,Monson3,Levesque,Wei1}. 
Moreover, eq.~(\ref{criticalT}) has 
only one solution for $T_c$. Numerical solution of the complete MFA equation 
for $T_c$ gives also only one critical point for small $\Delta$.

In the case of strong heterogeneity, the occupation of regions with
different binding energies occurs at quite different thermodynamic conditions
({\it e.g.}, chemical potentials or pressures). 
At low chemical potential, the fluid fills essentially the region with lower 
external potential {\it i.e.}, $\tau_1 << \tau_0$ (in the case of a large 
negative $\Delta$). Neglecting $\tau_1$ in  eq.~(\ref{bcoverage}), we obtain 
\begin{equation} 
\mu=-\epsilon_0-qW\theta+\ln 
\left[\frac{\theta}{c_0-\theta}\right]. 
\end{equation}
The criticality conditions (\ref{conditions}) lead to
$$
\theta_c=c_0/2; \qquad T_c/T_c^{bulk}=c_0.
$$
When the region of lower energy is completely filled,  $\tau_0=1$, the fluid 
density given in eq.~(\ref{bcoverage}) can be approximated as 
$\theta=c_0+(1-c_0)\tau_1$. This leads to
\begin{equation}
\mu=-\epsilon_0-qW\theta-\Delta+ kT\ln \left[\frac{\theta-c_0}{1-\theta}\right].
\end{equation}
The second critical point is located at
$$
\theta_c=(1+c_0)/2; \qquad T_c/T_c^{bulk}=1-c_0.
$$
Note that $c_0$ is the concentration of the sites with lower external 
potential and the change
$\Delta \to -\Delta$ is equivalent to $c_0 \to 1-c_0$.
The above approximate analysis reveals that there is only one critical point
in the case of weak heterogeneity while two critical points in strongly
heterogeneous systems.

As an illustration, we show in Figure~1 the spinodal curves (as a
function of fluid density $\theta$) for
different values of $\Delta$. The critical point is shifted to lower 
temperatures and
lower densities when $|\Delta|$ is increased. The picture is symmetrical
with respect to  $\theta=1/2$ for $\Delta>0$, this is equivalent to
switching between the regions of higher and lower external potentials. 
The high-temperature critical point corresponds to the condensation in the 
the larger sublattice ({\it i.e.}, that with a higher concentration). The 
matrix disorder plays only a marginal role, because essentially the same 
shift of the critical parameters is observed for fluids confined in regular 
geometries\cite{pore}. The second (low-temperature) critical point appears 
only when the heterogeneity exceeds some threshold, {\it i.e.}, 
$|\Delta|>|\Delta^*|$. The other familiar features, {\it i.e.},
lowering $T_c$ (for both transitions) and the shrinking of the density gap 
between the coexisting phases, are
also present. The inset displays the variation of the spinodals with the
concentration $c_0$ (in the case of two transitions, $|\Delta|>|\Delta^*|$). 

Figure~2 shows the variation of the critical densities with $|\Delta|$ 
(for $\Delta < 0$) for a given $c_0$ ($c_0=0.3$). This figure shows clearly 
that the threshold, $|\Delta^*|$,  for this case is near $4qW$.
It is quite natural that the threshold, $|\Delta^*|$, changes with $c_0$. 
The inset gives a "phase diagram" in the parameter space of $|\Delta^*|$ 
and $c_0$ which delimits the region of one or two phase transitions. This 
phase diagram shows that the lowest threshold is for the case of $c_0=c_1$ 
and that $|\Delta^*|$ becomes higher and higher when $|c_0-c_1|$ is 
increased. This variation can be interpreted as follows. When $|c_0-c_1| \to 
1$  (either $c_0 \to 1$ or $c_1 \to 1$), the system becomes spatially more 
and more homogeneous. In this case, the heterogeneity level, 
sufficient for the double transition,  can be reached only 
with a sufficiently high value of $|\Delta^*|$.  

In this subsection, we have shown that the bimodal random binding
lattice gas model 
is capable of reproducing the basic features of the phase diagram for a 
fluid confined in a random porous medium. The model ignores the role of 
porosity but focuses on the role of the disorder in the adsorbate binding 
energies. The second critical point is shown to survive only if the 
heterogeneity exceeds some threshold (measured by $|\Delta^*|$).

%%%%%%%%%%%%%%%%%%%%%%%%%%%%%
\subsection{Lattice gas with quenched impurities}
%%%%%%%%%%%%%%%%%%%%%%%%%%%%%

In order to take into account the matrix porosity and
the variation of local fluid coordination, we consider now the
lattice gas with quenched impurities\cite{Kierlik1,Kierlik2,Shapir2,Sarkisov} which is described by the following Hamiltonian,
\begin{equation}
H=-W\sum_{<ij>}t_it_jx_ix_j
-K\sum_{<ij>}[t_ix_i(1-x_j)+t_jx_j(1-x_i)]
\end{equation}
The sums are over nearest neighbors. For this reason $\sum_j(1-x_j)=p_i$ is
the number of the matrix nearest neighbors around the site $i$ and
$\sum_j x_j=q-p_i$.
MFA (see eq.~(\ref{MFA})) leads to the following simplified Hamiltonian,
\begin{equation}
\label{Hamil2}
H=\frac{W\theta^2}{2}[N_f
q-{\sum_i}' p_i]-qW\theta {\sum_i}' t_i
-(K-W\theta) {\sum_i}' t_ip_i.
\end{equation}
Note that in eq.~(\ref{Hamil2}) we count only the sites accessible to the fluid and
$\sum_i'=\sum_{i=1}^{N_f}$ denotes the summation over these sites.
The grand partition function for a particular matrix realization is given by
\begin{equation}
\Xi(\{p_i\})=
e^{-\beta W\theta^2
[qN_f-\sum_i' p_i]/2}
\prod_{i=1}^{N_f} \left[
1+e^{\beta[\mu+qW\theta+(K-W\theta)p_i]}\right]
\end{equation}
For the lattice model considered here, the distribution of matrix nearest neighbors around a fluid site can be written as,
\begin{equation}
f(p_i)=\sum_a c_a \delta_{p_i,p_a}, \qquad \sum_a c_a=1.
\end{equation}
Taking the average over $p_i$, we obtain the following expression for 
the pressure,
\begin{equation}
\label{pressure1}
P=\frac{kT}{N_f} \sum_{p_i} f(p_i) \ln\Xi(\{p_i\})=
-\frac{W\theta^2}{2}[q-{\overline p}]
+ kT\sum_a c_a \ln
\left \{
1+e^{\beta[\mu+qW\theta+(K-W\theta)p_a]} \right \},
\end{equation}
where ${\overline p}$ is the average nearest matrix neighbors around a site not occupied by a matrix particle, {\it i.e.},
\begin{equation}
{\overline p}=\sum_{p_i}f(p_i)\frac{1}{N_f} {\sum_i}' p_i=\sum_a c_a p_a.
\end{equation}
The average porosity of the matrix is given by
$\phi=1-{\overline p}/q$. From eq.~(\ref{pressure1}), it can be clearly 
seen that $P$ depends not only on ${\overline p}$, but also on the individual
${p_a}'s$. Therefore, two systems with the same average porosity
can behave differently.

The fluid density after the averaging over the matrix disorder is given by,
\begin{equation}
\label{coverage2}
\theta=
\sum_a c_a
\frac{e^{\beta[\mu+(q-p_a)W\theta+Kp_a]}}
{1+e^{\beta[\mu+(q-p_a)W\theta+Kp_a]}}.
\end{equation}
The result given in eq.~(\ref{coverage2}) agrees with that obtained by 
Sarkisov and Monson\cite{Sarkisov}. The only difference is that we do not 
work with local fluid densities. If the MFA (\ref{MFA}) is applied locally, 
{\it i.e.}, $t_it_j=t_i\theta_j+t_j\theta_i-\theta_i\theta_j$, we can 
recover the results of ref.~\cite{Sarkisov}. It can be readily checked that 
the pressure given in eq.~(\ref{pressure1}) and the density given in 
eq.~(\ref{coverage2}) satisfy Gibbs-Duhem equation, eq.~(\ref{GibbsDuhem}).  

From  eq.~(\ref{coverage2}), we can
see why the critical temperature for a confined fluid is lower
than in the bulk. The main reason is that the effective fluid-fluid
interaction inside the matrix is weaker because of the decreasing fluid
coordination number $q-p_a$. Eq.~(\ref{coverage2}) has quite the same structure
as eq.~(\ref{coverage}). Therefore, we can anticipate that the results to be obtained with the model considered in this subsection will be very similar to those in the previous subsection. It is expected that there exist a region in the parameter space of 
$p_a$'s (local number of matrix neighbors), $c_a$'s (concentrations) and $K$ 
(fluid-matrix interaction) in which the double criticality appears.
To confirm this, we carried out numerical calculations again for a bimodal case with two
coordinations $p_0$ and $p_1=p_0+\Delta$, such that the average number of matrix
neighbors is,
\begin{equation}
{\overline p}=c_0p_0+(1-c_0)p_1= p_0+(1-c_0)\Delta.
\end{equation}
Then, eq.~(\ref{coverage2}) gives a typical bimodal result with
two sublattices. As in the previous subsection, we can see either one or two phase 
transitions depending on the specific
values of $c_0$ and $\Delta$. The spinodal curve with the double criticality is presented in
Figure~3 which shows all the characteristic features we have seen already (lowered
$T_c$'s and narrowed coexistence gap). The spinodal curve is again
symmetrical with respect to $\theta=1/2$ when the sign of the
fluid-matrix interaction, $K$, is changed. However, this does not change the critical
temperature (for both transitions). Recall that the positive $K$ describes a repulsive fluid-matrix interaction. In this case, it is
difficult to associate one of the two phase transitions with the wetting phenomenon.

No double criticality has been found from the rigorous results \cite{Sok} for this model 
on a Bethe lattice. Nevertheless, it should be noted
that the effective field distribution found by the authors of \cite{Sok} is 
multimodal
and temperature-dependent (see figures 3,4 in \cite{Sok}). The distinction
between the main peaks becomes less clear-cut with increasing temperature and
site dilution. We may suppose that the range of parameters explored
was out of the threshold, necessary for observing the double criticality.
Another important point is that the authors of \cite{Sok} have studied
the case of fixed field, $h=\mu/2+qW/4$, which is not the case for our study 
here. Such a
constraint fixes a given thermodynamic path (pressure, temperature) along
which the state of the fluid is varied. As a result, their field distribution changes its
degree of asymmetry with the temperature. It is known that the degree of
asymmetry is crucial in observing the double
criticality in the RFIM language \cite{RFIM}.

In this subsection, we studied a model which takes into account  
the presence of matrix particles. The explicit description of matrix 
particles introduces two key ingredients which play an important role in 
the fluid phase behavior, {\it i.e.}, the local fluid 
coordination and the fluid-matrix interaction. It is remarkable that the 
mean-field result of this model is very similar to that obtained in the 
previous subsection for a model which only embodies the spatial 
heterogeneity. So, it is tempting and also looks plausible to consider the 
space heterogeneity as the key prerequisite of the double criticality. The 
real appearance of the double criticality is determined by the strength of 
the heterogeneity which is described by $\Delta^*$ for both models 
considered above. Although the numerical values of $\Delta^*$ differs for 
the two models, the main physics is not altered at all. The space 
heterogeneity can have different origins. For the model considered in this 
subsection, the double criticality is related to the coexistence of two 
polymorph liquid phases with different coordination numbers. This 
polymorphism is induced by the heterogeneity in the number of matrix nearest 
neighbors.

%%%%%%%%%%%%%%%%%%%%%%%%%%

\section{Role of effective distribution}

%%%%%%%%%%%%%%%%%%%%%%%%%%
In the previous section, we have seen that a bimodal matrix
distribution may lead either to one or to two critical points. 
Now, a subtle question one may ask is whether it is possible to observe 
two phase transitions when the matrix distribution is apparently unimodal? 
Before answering this question, it is useful to discuss how the pore size 
distribution is usually determined. Even for rigid matrices, such a 
determination is not a straightforward task. For the models frequently used 
in computer simulations ({\it e.g.}, matrix of hard spheres), the 
distribution of matrix particles is not related directly to a pore size 
distribution. The latter is usually determined through the 
tessellation\cite{Wei1} of the void space. It is to be noted that such 
tessellation methods do not allow for identifying closed cavities. The 
closed pores are inaccessible to the fluid in real experiments or in  
simulations which mimic the real fluid adsorption. 
In these cases, the matrix distribution seen by the fluid is 
different from that determined without excluding the closed pores.  Recent 
studies \cite{Sok,Gubbins} have demonstrated the importance of the pore 
blocking and connectivity effects. 

In order to analyze the role of closed cavities, one can
assume that the nominal pore size
distribution $P(r)$ ({\it i.e.}, that including the contribution of 
closed pores) is gaussian with the mean pore size $\sigma_2$ and dispersion 
$\delta_2$. Guassian distributions were indeed found with the tessellation 
method for the matrix samples used in simulations\cite{Wei1}. Since the 
closed pores are generated in the same way as the open ones, it can be 
reasonably expected that the distribution of the closed pores, $p(r)$, is 
similar to the nominal one but characterized by $\sigma_1$ and $\delta_1$. 
If the fluid accessible pores and the closed cavities are uncorrelated, 
then a non-normalized distribution of fluid accessible pores $\varrho(r)$ is 
given by the product of the nominal distribution and the probability of 
finding an open pore ($[1-p(r)]$), {\it i.e.}, 
$$\varrho(r)=P(r)\left[1-p(r)\right].$$ 
Even when $P(r)$ and $p(r)$ are unimodal distributions, 
$\varrho(r)$ could be bimodal. Such a situation is shown in Figure~4. 
One can see that the effective pore size distribution, $\varrho(r)$, 
becomes bimodal when there is a significant overlap between the peak of 
$P(r)$ and that of $p(r)$. Just as the appropriate combination of $\Delta$ 
and $c_0$ leads to the double criticality,  the appropriate combination of 
$\sigma_1$,$\sigma_2$, $\delta_1$ and $\delta_2$ results in an effective 
bimodal pore size distribution which can lead to the double criticality. 
This might be considered as one of possible explanations of the puzzling 
results found in the computer simulation \cite{Wei1}.

From a more general point of view, the pore size distribution can also depend 
on the fluid state because of the pore blocking effects\cite{Gubbins} or due 
to the matrix reaction to the fluid adsorption. This is evident for real 
world experiments with adsorption in high-porosity materials like 
aerogels\cite{aerogel1,aerogel}. Upon adsorption, such matrices can contract 
or expand in volume and thus change their pore size distribution. This 
effect is not exclusive to relatively soft gel-like matrices but also 
quite common for various intercalation compounds\cite{PRB,JPCB}.
In fact, the distribution of the pore space effectively seen by a fluid  
depends on both the matrix preparation and the 
fluid state. In our language, this is equivalent to say that the distribution 
is conditional, $P(\epsilon)=P(\epsilon|\theta)$, which is conditioned 
by the fluid state characterized here by the fluid density, $\theta$. For 
simplicity, we consider explicitly the binding energy distribution but all 
the conclusions can be easily translated into the porosity language.

As we have seen in the previous sections, the average fluid density can be 
obtained from that for a particular disorder realization, 
$\theta(\epsilon)$, through the following relation, 
\begin{equation}
\theta=\int d\epsilon P(\epsilon|\theta) \theta(\epsilon).
\end{equation}
Now the compressibility becomes
\begin{equation}
\label{compress}
\chi=\frac{1}{\theta^2}\frac{\partial \theta}{\partial \mu}=
\frac{1}{\theta^2}\frac{\partial}{\partial \mu}
\int d\epsilon P(\epsilon|\theta) \theta(\epsilon)=\frac{1}{\theta^2}\left [
\int d\epsilon P(\epsilon|\theta)
\frac{\partial \theta(\epsilon)}{\partial \mu}
+
\int d\epsilon \frac{\partial P(\epsilon|\theta)}{\partial \theta}
\frac{\partial \theta}{\partial \mu}\theta(\epsilon) \right ]. 
\end{equation}
Here, the first term is an average over disorder and
the second term takes into account the matrix reaction {\it i.e.}, the
modification of the matrix distribution induced by adsorbates.
Eq.~(\ref{compress}) can be rewritten as
\begin{equation}
\chi=\chi_0 \left[1-\int d\epsilon \frac{\partial P(\epsilon|\theta)}
{\partial \theta} \theta(\epsilon)
\right]^{-1}, 
\end{equation}
where
\begin{equation}
\chi_0=\frac{1}{\theta^2}\int d\epsilon P(\epsilon|\theta)
\frac{\partial \theta(\epsilon)}{\partial \mu}.
\end{equation}
In the case of a unimodal $P(\epsilon|\theta)$ (no double
criticality), a divergent $\chi_0$ (at some density $\theta=\theta_0$)
corresponds to a phase transition resulting mainly from the intrinsic fluid
properties. The matrix only rescales the critical parameters and the 
behavior of the confined fluid is somehow bulk-like. But, under the condition
\begin{equation}
\int d\epsilon \frac{\partial P(\epsilon|\theta)}
{\partial \theta} \theta(\epsilon) =1,
\end{equation}
$\chi$ is again singular (at another fluid density $\theta=\theta_1$). 
We observe under this condition a second transition. This effect
is quite similar to the polymorphism in simple fluids with density-dependent
pair potentials\cite{Baus,Tejero}. Moreover,
due to the matrix reaction, we can see the criticality even if the bulk-like
behavior is not critical ({\it i.e.}, $\chi_0$ being finite). 
This offers a possibility for creating a critical state under 
conditions, different from the bulk critical point. This 
possibility might find interesting technological applications. 

We would like to emphasize that in fact the pore size distributions measured 
experimentally  are the conditional ones. From the usual techniques,
the pore size distribution is often deduced by fitting adsorption 
isotherms\cite{aerogel1}. It is well known that the results of such 
measurements depend on the fluid state and on the used probe molecules. From 
the theoretical point of view, such a distribution naturally appears when 
the maximum entropy approach \cite{SUSCINF} is applied to the determination 
of the relevant distribution (for $\epsilon$ in the case considered here) 
from some indirect data $\theta={\overline {\theta(\epsilon)}}$ (the 
over-bar denotes the average over disorder).

In this section, we described a scenario in which a nominal unimodal matrix
distribution, found through analyzing the matrix geometry, can behave as an
effectively bimodal if the accessibility of the pores to the fluid is 
taken into account. Alternatively, one can observe the double criticality 
even for a unimodal distribution, provided that it is sensitive to the fluid 
thermodynamic state. In this case, one critical point corresponds to a 
rescaled bulk-like transition, while the other transition is related to the 
matrix reaction to the presence of adsobates.
%%%%%%%%%%%%%%%%%%%%%%%%

\section{Conclusion}

%%%%%%%%%%%%%%%%%%%%%%%%

The main question studied in the present work is how the heterogeneity 
of a hosting system is related to the occurrence
of two critical points for a fluid adsorbed in it. 
For answering this question, we analyzed in details the role of several 
relevant factors. Our analysis is based on an analytically tractable lattice 
model with a bimodal distribution of binding energies or local porosities. 
We find that the the bimodality itself does not guarantee the existence of 
two critical points. The adsorbed fluid can undergo two phase transitions 
only when the host heterogeneity is sufficiently 
strong. For the models considered here, this is manifested through 
the existence of a threshold, $\Delta^*$,  for the 
appearance of the double criticality. For two samples having (due to the 
preparation) the same mean binding energy $\overline 
\epsilon=\epsilon_0+(1-c_0)\Delta$ or mean porosity 
$\phi=1-[p_0+(1-c_0)\Delta]/q$, one can observe either one or two critical 
points depending on the particular values taken by $c_0$ and $\Delta$. This 
provides the criterion for predicting the occurrence of the double 
criticality for the lattice models considered here. Nevertheless, it does 
not seem to be straightforward to translate the criterion obtained here to 
the case of off-lattice models.

Within the framework of the models studied here, the two phase
transitions correspond to the condensation in the regions of different
binding energy or local porosity.  In the latter case,
the nature of the
double criticality is related to the coexistence of two polymorph liquid
phases with different coordination numbers. This polymorphism is induced by
the heterogeneity in the number of matrix nearest neighbors. The reduced
fluid coordination number (compared to the bulk) leads to the
well-known effects: lowering of $T_c$ and shrinking of the coexistence density
gap.

We argue that in order to give an appropriate analysis of simulation results 
in terms of a pore size distribution function,  one
should be aware that the matrix distribution seen effectively by the 
adsorbed fluid can depend on many factors, {\it e.g.}, presence of closed 
pores, non rigidity of the matrix etc.. A nominal unimodal matrix
distribution, found through analyzing the matrix geometry, can behave as a
bimodal in the presence of a fluid if closed pores exist. This could be a
possible explanation of recent puzzling simulation results\cite{Wei1}.
Alternatively,
one can observe the double criticality even in the case of an unimodal 
distribution, provided that it is sensitive to the fluid thermodynamic 
state. In this scenario, one critical point corresponds to a rescaled 
bulk-like transition, while the other transition is induced by the matrix 
reaction to the presence of the fluid. This double criticality mechanism is 
quite similar to the polymorphism in fluids with density-dependent 
interactions\cite{Baus,Tejero}.

Finally, we can wonder also why no more than two phase transitions 
have been observed from simulations up to now. In principle, we can observe 
three critical points with the models studied here when a trimodal 
distribution is considered with $c_0,c_1,c_2$ and 
$\epsilon_0,\epsilon_1,\epsilon_2$. In the light of the results obtained in 
this work, we expect to observe three phase transitions only under the 
conditions that $\Delta_1=\epsilon_0-\epsilon_1$ and 
$\Delta_2=\epsilon_1-\epsilon_2$ are different enough. The same argument 
applies to the matrix coordination differences. But in this case, the above 
conditions can not be fulfilled easily because neither $\Delta_1$ 
($p_0-p_1$) nor $\Delta_2$ ($p_1-p_2$) can exceed the maximal number of 
nearest matrix neighbors, $q$, which is restricted by the dimensionality of 
space and by steric effects. This makes the region of three phase 
transitions  very narrow (even if it exists in the parameter phase 
diagram). This is probably why till now only two phase transitions were 
found.

%%%%%%%%%%%%%%%%%%%%%%%%%%%

%%%%%%%%%%%%%%%%%%%%%%%%%%%%%%%%

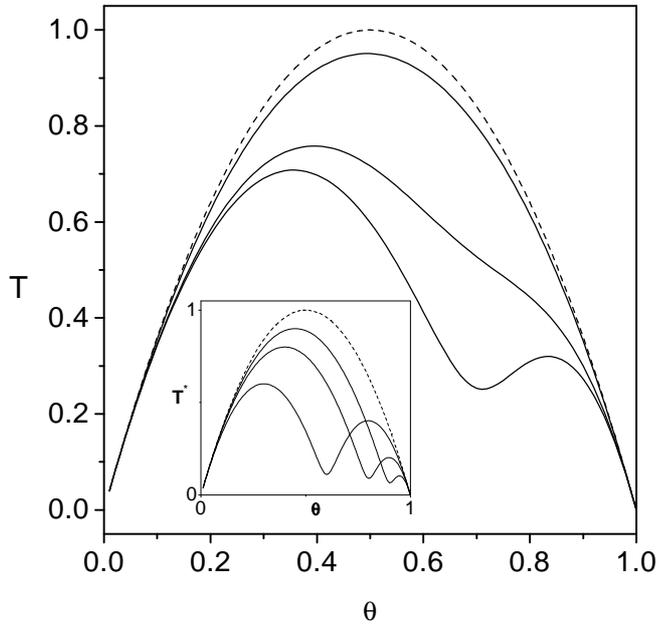
\begin{figure}
\caption{Spinodal curves for the random binding model. From the top to the
bottom $\Delta=0,-1,-3,-5$, q=4, c=0.7. The inset: $c=1,0.9,0.8,0.6$-from
the top to the bottom, $\Delta=-7$. The temperature is dimensionless
$T=T^*/T_{c,b}^*$}
\end{figure}

%%%%%%%%%%%%%%%
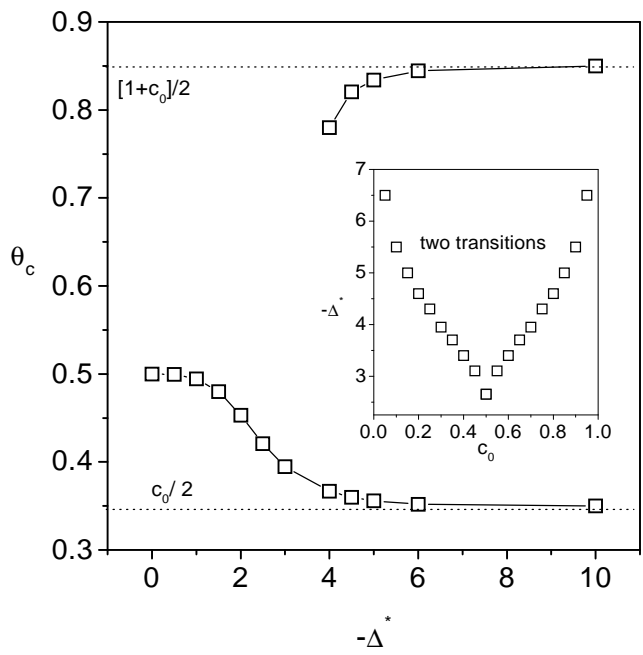
\begin{figure}
\caption{The critical coverage $\theta_c$ as a function of the energy
difference $\Delta^*=\Delta/qW$ at $c_0=0.7$. The inset
demonstrates how the threshold varies with the concentration $c_0$.
Horizontal dotted lines correspond to our analytical estimation
for large $\Delta$ (see text).
}
\end{figure}
%%%%%%%%%%%%%%%%%%%%
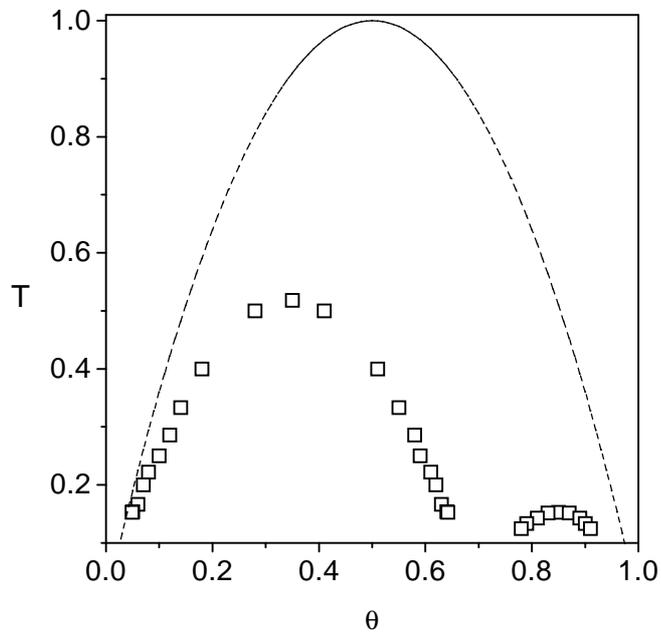
\begin{figure}
\caption{Spinodal curves for the model with quenched impurities. The dashed
curve corresponds to the bulk case ($p_0=0$, $\Delta=0$), the symbols -
$K/qW=10$, $c=0.4$, $p_0=2$, $q=4$, $\Delta=-1$. 
}
\end{figure}
%%%%%%%%%%%%%%%%%%%%%%%%%%%
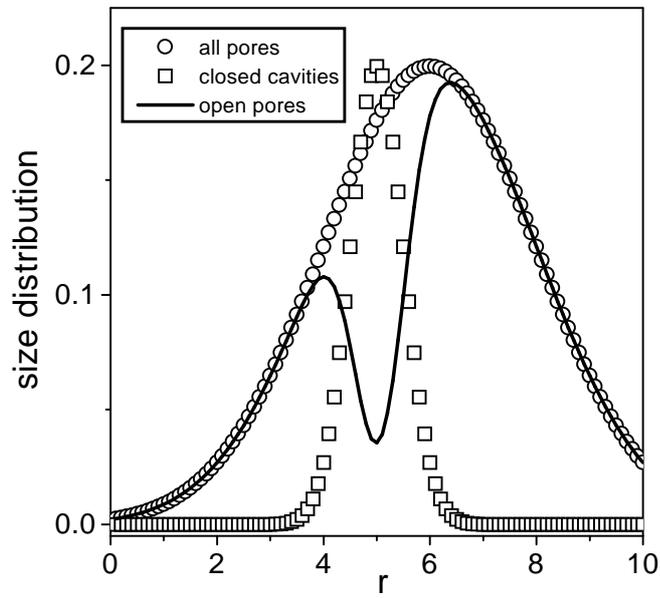
\begin{figure}
\caption{The overall pore size distribution $P(r)$, the distribution
of closed cavities $p(r)$, and the resulting accessible pore distribution
$\varrho(r)$. Both  $P(r)$ and $p(r)$ are gaussian with $\delta_2=2,
\sigma_2=6$ and $\delta_1=0.5, \sigma_1=5$ (see text). For pictorial
purposes $p(r)$ (squares) is reduced by a factor of $1/4$.}
\end{figure}
%%%%%%%%%%%%%%%%%%%%%%%

\begin{thebibliography}{99}
\bibitem{comment}
The literature on this field is quite extensive. We are not going to
present an exhaustive list, but just mention the works that are closely
related to our subject.
\end{thebibliography}
\end{document}